\documentclass[sigconf, screen, authorversion]{acmart}

\AtBeginDocument{%
 }

\copyrightyear{2025}
\acmYear{2025}
\setcopyright{rightsretained}
\acmConference[SIGCSE TS 2025]{Proceedings of the 56th ACM Technical Symposium on Computer Science Education V. 1}{February 26-March 1, 2025}{Pittsburgh, PA, USA}
\acmBooktitle{Proceedings of the 56th ACM Technical Symposium on Computer Science Education V. 1 (SIGCSE TS 2025), February 26-March 1, 2025, Pittsburgh, PA, USA}
\acmDOI{10.1145/3641554.3701974}
\acmISBN{979-8-4007-0531-1/25/02}

\makeatletter
\gdef\@copyrightpermission{
  \begin{minipage}{0.3\columnwidth}
   \href{https://creativecommons.org/licenses/by/4.0/}{\includegraphics[width=0.90\textwidth]{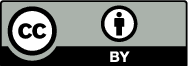}}
  \end{minipage}\hfill
  \begin{minipage}{0.7\columnwidth}
   \href{https://creativecommons.org/licenses/by/4.0/}{This work is licensed under a Creative Commons Attribution International 4.0 License.}
  \end{minipage}
  \vspace{5pt}
}
\makeatother

\usepackage{balance} 
\usepackage[utf8]{inputenc} 
\usepackage{colortbl}
\usepackage{subcaption}
\usepackage{diagbox}
\usepackage{multirow}
\usepackage{xspace}
\usepackage{enumitem} 
\usepackage{mdframed} 
\usepackage{caption}
\usepackage{subcaption}
\usepackage{wrapfig}
\usepackage{etoolbox}
\usepackage{verbatim}
\usepackage{amsmath}
\usepackage{graphicx}
\usepackage{tabularray} 
\usepackage{caption}
\usepackage{multirow}
\usepackage{makecell}
\usepackage{bbm}

\usepackage{hyperref}
\hypersetup{ %
	pdfauthor={},
	pdftitle={},
	pdfsubject={},
    pdflang={en},
	bookmarksnumbered,
	pageanchor=true,
	plainpages=false,
	colorlinks=true,%
	linktocpage=true,
	citecolor=MidnightBlue,%
	filecolor=BrickRed,%
	linkcolor=NavyBlue,%
	urlcolor=NavyBlue 
}

\usepackage{environ}
\usepackage{esvect}
\usepackage[ruled,vlined]{algorithm2e}
\usepackage{algcompatible}
\usepackage[noend]{algpseudocode}

\NewEnviron{boxcode2col}[4]{
    \begin{center}
        \scalebox{#3}{
            \setlength\tabcolsep{4pt}
            \renewcommand{\arraystretch}{#4}
            \begin{tabular}{|p{#1}p{#2}|}   
                \hline
                \BODY
                [1ex]
                \hline
            \end{tabular}
        }
    \end{center}
}

\NewEnviron{boxcode}[3]{
    \begin{center}
        \scalebox{#2}{
            \setlength\tabcolsep{4pt}
            \renewcommand{\arraystretch}{#3}
            \begin{tabular}{|p{#1}|}
                \hline
                \vspace{0.1mm}    
                \BODY
                \\\hline
    		\end{tabular}
    	}
    \end{center}
}

\usepackage{graphicx}

\newcommand*{\img}[1]{%
    \raisebox{-.2\baselineskip}{%
        \includegraphics[
        height=0.9\baselineskip,
        width=\baselineskip,
        keepaspectratio,
        ]{#1}%
    }%
}

\newcommand{\textcode}[1]{{\fontfamily{cmtt}\selectfont #1}\xspace}

\usepackage{listings}

\definecolor{mGreen}{rgb}{0,0.6,0}
\definecolor{mGray}{rgb}{0.5,0.5,0.5}
\definecolor{mPurple}{rgb}{0.58,0,0.82}
\definecolor{RoyalPurple}{RGB}{120, 81, 169}
\definecolor{OliveGreen}{RGB}{128, 128, 0}

\lstdefinestyle{CStyle}{
    backgroundcolor=\color{white},   
    commentstyle=\color{mGreen},
    keywordstyle=\color{blue},
    numberstyle=\tiny\color{mGray},
    stringstyle=\color{mGreen},
    basicstyle=\footnotesize,
    breakatwhitespace=false,         
    breaklines=true,                 
    captionpos=b,                    
    keepspaces=true,                 
    numbers=left,                    
    numbersep=5pt,                  
    showspaces=false,                
    showstringspaces=false,
    showtabs=false,                  
    tabsize=2,
    language=C
}

\newcommand{\TechChatGPT}{\ensuremath{\textsc{GPT3.5}}}
\newcommand{\TechGPTFour}{\ensuremath{\textsc{GPT4o}}}

\newcommand{\buggy}{\ensuremath{\text{\textcode{C}}_\text{B}}}
\newcommand{\fixed}{\ensuremath{\text{\textcode{C}}_\text{F}}}
\newcommand{\problem}{{\text{\textcode{P}}}}

\begin{document}

\title[BugSpotter: Automated Generation of Code Debugging Exercises
]{BugSpotter: Automated Generation of Code Debugging Exercises
}

\author{Victor-Alexandru P{\u a}durean}
\affiliation{
  \institution{MPI-SWS}
  \city{Saarbr{\"u}cken}
  \country{Germany}  
}
\email{vpadurea@mpi-sws.org}

\author{Paul Denny}
\affiliation{
  \institution{University of Auckland}
  \city{Auckland}
  \country{New Zealand}  
}
\email{paul@cs.auckland.ac.nz}

\author{Adish Singla}
\affiliation{
  \institution{MPI-SWS}
  \city{Saarbr{\"u}cken}
  \country{Germany}  
}
\email{adishs@mpi-sws.org}



\begin{abstract}
Debugging is an essential skill when learning to program, yet its instruction and emphasis often vary widely across introductory courses.  In the era of code-generating large language models (LLMs), the ability for students to reason about code and identify errors is increasingly important.  However, students frequently resort to trial-and-error methods to resolve bugs without fully understanding the underlying issues. Developing the ability to identify and hypothesize the cause of bugs is crucial but can be time-consuming to teach effectively through traditional means. This paper introduces BugSpotter, an innovative tool that leverages an LLM to generate buggy code from a problem description and verify the synthesized bugs via a test suite.  Students interact with BugSpotter by designing \emph{failing test cases}, where the buggy code's output differs from the expected result as defined by the problem specification. This not only provides opportunities for students to enhance their debugging skills, but also to practice reading and understanding problem specifications. We deployed BugSpotter in a large classroom setting and compared the debugging exercises it generated to exercises hand-crafted by an instructor for the same problems. We found that the LLM-generated exercises produced by BugSpotter varied in difficulty and were well-matched to the problem specifications. Importantly, the LLM-generated exercises were comparable to those manually created by instructors with respect to student performance, suggesting that BugSpotter could be an effective and efficient aid for learning debugging.
\end{abstract}





\maketitle


\section{Introduction}\label{sec.introduction}

Debugging is an essential skill for programming, yet there is little consistency in how it is taught \cite{whalley2021novice}.
A landmark review by McCauley et al. covered various educational perspectives on debugging, highlighting that it is both difficult for novices to learn and challenging for computer science educators to teach \cite{mccauley2008debugging}.  Similar work has explored common difficulties faced by students when learning debugging \cite{fitzgerald2008debugging}.  For example, Whalley et al. observed that novice programmers often employ vague and imprecise methods for hypothesis generation and verification, relying heavily on guesswork rather than using a systematic approach \cite{whalley2021novice}.

\looseness-1With large language models (LLMs) capable of automatically generating code, the ability for students to identify and debug errors becomes even more critical \cite{denny2024cacm}.  Educators must ensure that students are equipped with the skills to critically evaluate and correct LLM-generated code.  Traditional debugging tasks typically involve presenting students with buggy code (i.e., code containing bugs) and asking them to identify and fix the issues.  Frameworks for teaching debugging have also been proposed, providing more structured approaches to enhance debugging instruction \cite{li2019towards}. Recent work by Ma et al. also highlights the importance of training students to hypothesize the causes of code errors, encouraging them to develop and test hypotheses about code defects systematically \cite{Ma_2024}.

Creating a diverse range of debugging exercises that offer automated feedback could make debugging instruction more consistent and frequent.  In this paper, we present a new tool called BugSpotter which is designed to generate debugging exercises using LLMs. BugSpotter generates buggy code from problem descriptions and verifies these bugs with a test suite. Students interact with the tool by designing failing test cases, which not only helps them practice debugging but also improves their ability to read and understand problem specifications. This method aligns with work on metacognitive scaffolding which has demonstrated the value of test case generation for helping with problem understanding \cite{denny2019closer}.  

\begin{figure*}[t!]
\centering
	\begin{minipage}{\linewidth}
    {
        \begin{subfigure}{0.48\linewidth}
        {\centering
            \includegraphics[width=\linewidth]{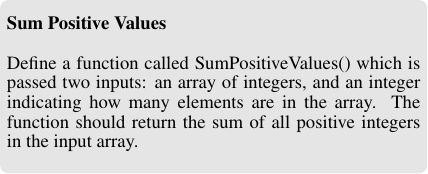}
            \vspace{-5.5mm}
            \caption{Problem specification}				
            \label{fig.illustration_p1.description}
        }
        \end{subfigure}
        \ \ \ \ \ \ \
        \begin{subfigure}{0.44\linewidth}
        {
            \setlength{\fboxsep}{2pt}\fbox{\includegraphics[width=\linewidth]{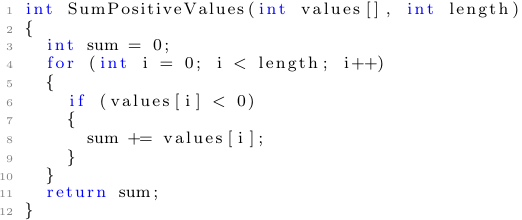}}
            \vspace{-4.5mm}
            \caption{Buggy code}				
            \label{fig.illustration_p1.code}
        }
        \end{subfigure}
    }
    \end{minipage}
    \begin{minipage}{\linewidth}
    {\ \
        \begin{subfigure}{0.95\linewidth}
        {\centering
            \vspace{0.1cm}
            \setlength{\fboxsep}{-2pt}\fbox{
            \includegraphics[width=\linewidth, trim=0mm 0.7mm 0mm 0.1mm]{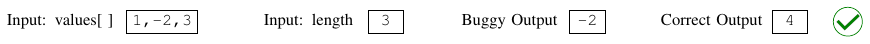}
            }
            \vspace{-4.5mm}
            \caption{Test case design}				
            \label{fig.illustration_p1.testcase}
        }
        \end{subfigure}
    }
    \end{minipage}
    \vspace{-4mm}
     \caption{
     \looseness-1Illustration of a debugging exercise from BugSpotter for Problem 1, where a student's objective is to design a failing test case.  \textbf{(a)} shows the problem specification. \textbf{(b)} shows the buggy code. \textbf{(c)} shows a successfully designed failing test case. A failing test case is composed of the function's input, its buggy output, and the correct output, according to the problem specification. In this case, the student successfully solved the exercise (tick \img{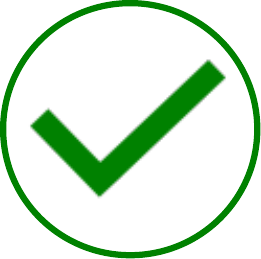}) by (1) providing a test case that leads to the buggy code generating a different output to the correct code, (2) providing the buggy output that matches what is obtained by executing the buggy code on the input, and (3) providing the correct output that matches what is obtained by executing the reference solution code on the input.
     }
     \vspace{-2mm}
    \label{fig.illustration_p1}
\end{figure*}

Figure~\ref{fig.illustration_p1} illustrates a debugging exercise generated by BugSpotter, with more information about how these exercises are created in Section~\ref{sec.method}.  We deployed BugSpotter in a large classroom setting and conducted a comparative analysis between LLM-generated debugging exercises and those manually created by instructors.  We evaluate our experience of using BugSpotter by answering the following questions:
\setlength{\leftmargini}{2em}
\begin{itemize} 
    \item \textbf{RQ1:} How do experts evaluate the quality of the debugging exercises generated by BugSpotter?
    \item \textbf{RQ2:} How does the difficulty of the generated exercises vary as measured by students’ performance on the exercises and expert-classified difficulty?
    \item \textbf{RQ3:} How does student success on LLM-generated debugging exercises compare to that on instructor-designed exercises?
\end{itemize}


\section{Related Work}\label{sec.relatedwork}
\textbf{The importance of debugging.}
Previous works have emphasized the importance of students acquiring debugging skills \cite{DBLP:conf/latice/ChenWL13, DBLP:conf/sigcse/KafaiDFLC19, whalley2021novice}. There are various strategies that students can use \cite{fitzgerald2008debugging,DBLP:conf/sigcse/MurphyLMSTZ08,DBLP:conf/ace/MacNeil0T0BHSK24}, yet they usually rely on guesswork \cite{whalley2021novice}. Explicitly teaching students debugging strategies has been demonstrated to be an effective method for promoting a more systematic approach \cite{DBLP:conf/sigcse/KoLHKKQAP19}. Despite this, both learning and teaching debugging remain challenging \cite{whalley2021novice}, thus novel frameworks for teaching debugging are necessary \cite{li2019towards}. Notably, the process of fixing the bug when its location is known is easily carried out \cite{fitzgerald2008debugging}, but locating the bug and understanding the functionality of the buggy code are considered difficult tasks \cite{mccauley2008debugging}. Our work focuses on automatically generating buggy codes for which students design failing test cases, practicing their skills for problem understanding, bug localization, and code comprehension.

\textbf{Reducing educator workload.}
\looseness-1Reducing the workload of educators is an ongoing endeavor in the education literature. One early initiative at curating a large reusable repository of multiple-choice questions for computer science is the Canterbury QuestionBank \cite{DBLP:conf/iticse/SandersACEGJLMP13}. Other approaches include student sourcing \cite{DBLP:conf/sigcse/DennyLTH11}, or designing tools to help educators create 
a range of programming-related
exercises \cite{DBLP:conf/iticse/LojoF22,DBLP:conf/sigcse/CaracoLVF24}. A recent line of research focuses on the automation of exercise creation, especially in introductory block-based visual programming domains \cite{DBLP:conf/nips/AhmedCEFGRS20,DBLP:conf/aied/GhoshTDS22,padurean2024neural,padurean24benchmarking}. Finally, in the era of LLMs, a suite of new tools leveraging generative models have been proposed \cite{DBLP:conf/icer/SarsaDH022,DBLP:journals/corr/abs-2306-17156,DBLP:conf/sigcse/Gutierrez0L24,DBLP:conf/sigcse/JordanLR24}. For example, HypoCompass \cite{Ma_2024} makes LLMs act as novices seeking help, placing students in the role of tutors. In contrast, BugSpotter generates debugging exercises focusing on the creation of failing test cases, allowing students to practice problem and code comprehension, along with bug localization.

\textbf{LLMs for programming education.}
\looseness-1Besides exercise creation, LLMs have been explored for other educational purposes \cite{DBLP:journals/corr/abs-2306-17156,DBLP:journals/corr/abs-2402-01580,molina2024leveragingllmtutoringsystems}, such as generating natural language feedback for students \cite{DBLP:conf/edm/PhungCGKMSS23,DBLP:conf/lak/PhungPS0CGSS24,DBLP:conf/sigcse/0001HSRDPB23,kotalwar2024hintsinbrowser}. Moreover, Nguyen et al. focus on generating buggy student attempts for block-based programming \cite{DBLP:journals/corr/abs-2310-10690}. Recently, the concept of prompt problems was proposed as a paradigm shift in this field \cite{DBLP:conf/sigcse/00010PLABR24,smith2024prompting}. BugSpotter uses LLMs to address a less-explored topic: the generation of debugging exercises.


\section{BugSpotter: Methodology and Tool}\label{sec.method}
\looseness-1The BugSpotter tool aims to help students read and understand problem specifications, reason about buggy code, and design test cases that reveal bug(s). When interacting with BugSpotter, a student is shown a code debugging exercise, comprising a problem specification and a buggy code for the problem. To solve this, the student has to design a test case that will reveal the error in the code (see Figure~\ref{fig.illustration_p1}). In the following, we introduce debugging exercises and then present our approach for generating such exercises using LLMs.

\subsection{Code Debugging Exercises}
We now describe the typical flow of a code debugging exercise generated with BugSpotter. Let us first define a problem \problem{}, a buggy code \buggy{}, and its fixed version \fixed{}. A problem \problem{} comprises a problem specification and a test suite. A buggy code \buggy{} fails at least one test case in \problem{}'s test suite. Its fixed version \fixed{} can be obtained by making small modifications to \buggy{} in order to pass \problem{}'s entire test suite. 

\looseness-1A student is presented with a specification of a problem \problem{} and corresponding buggy code \buggy{}. The problem specification requires implementing a single function, and \buggy{} represents a buggy version of this function. Figures~\ref{fig.illustration_p1.description}~and~\ref{fig.illustration_p1.code} show the specification for the ``Sum Positives'' problem along with an example buggy code \buggy{}. The student's objective is to provide a \emph{failing test case} that reveals why \buggy{} is incorrect. The test case should include an input to the function in \buggy{}, the incorrect output from the buggy function, and the expected correct output as defined by the problem specification. The student successfully solves the exercise if the following criteria are met: 
\setlength{\leftmargini}{2em}
\begin{itemize} 
\item[\textbf{(1)}] the outputs of \buggy{} and \fixed{} are different when run on the provided input (i.e., the test case is indeed a failing test case);
\item[\textbf{(2)}] the output of \fixed{} when run on the provided input matches the provided correct output;
\item[\textbf{(3)}] the output of \buggy{} when run on the provided input matches the provided buggy output.
\end{itemize}
\looseness-1Validation of the proposed failing test case is done automatically. Figure~\ref{fig.illustration_p1.testcase} shows an example where a student provides a test case meeting all required criteria. After completing the current exercise, the student can request another exercise. BugSpotter will then generate a new exercise based on the same problem, providing further opportunities for the student to identify different kinds of bugs. The process of generating a debugging exercise is detailed below.

\begin{figure}[t!]
\centering
	\includegraphics[width=\linewidth]{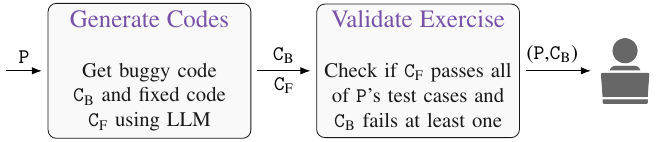}
    \vspace{-6mm}
     \caption{BugSpotter's exercise generation pipeline.
     }
     \vspace{-3mm}
    \label{fig.pipeline}
\end{figure}

\subsection{Exercise Generation Pipeline}
We now give details about BugSpotter's debugging exercise generation process, which comprises two stages: (a) code generation using LLMs and (b) validation of the exercise through execution. The generation process starts with a problem \problem{}, encompassing its problem specification and its test suite, as illustrated in Figure~\ref{fig.pipeline}.

\textbf{Code generation stage.}
\looseness-1We leverage an LLM for generating the buggy code. We provide the LLM with \problem{}'s problem specification, and prompt the LLM to reason about possible bugs that students may introduce while working on the problem's solution code. We then ask the LLM to generate a buggy code \buggy{}, its fixed version \fixed{}, and an explanation of the bug. The motivation to ask for reasoning, explanation, and fixed code \fixed{} draws inspiration from Chain-of-Thought \cite{DBLP:conf/nips/Wei0SBIXCLZ22}, encouraging the LLMs to carefully reason about the buggy code and how it differs from a correct code. Additionally, \fixed{} will also play a role during validation. We enforce JSON formatting for easily extracting the information from the LLM's response. Figure~\ref{fig.prompt} illustrates the prompt used in the tool for generating buggy codes.

\textbf{Exercise validation stage.}
\looseness-1As we focus on delivering semantic bugs well-matched to problem specifications, the validation stage is aimed at filtering out syntactic bugs and buggy codes unrelated to the problem. First, we filter out instances where \fixed{} or \buggy{} do not successfully compile. Second, we check whether \fixed{} can correctly solve the given problem using the test suite provided in \problem{}. The main intuition behind checking the correctness of \fixed{} is to filter out the instances where the LLM was inconsistent and could not generate a fixed version, potentially leading to confusing buggy code that does not match the given problem. Third, we run \buggy{} using the same test suite. For \buggy{} to be valid for the debugging exercise, it should fail at least one of the test cases in the test suite, either by raising a runtime error or by producing an output different from the expected one. Otherwise, \buggy{} would be a correct implementation and we will not use it as part of an exercise. Fourth, we filter out instances that run longer than a given time limit. Finally, if both \fixed{} and \buggy{} are successfully validated, we deliver the exercise as output.

\begin{figure}[t!]
\centering
	\begin{tabular}{||p{0.94\linewidth}||}
        \hline
        \multicolumn{1}{||c||}{\color{RoyalPurple}Prompt to Generate Buggy Codes} \\ 
        \small
        Below is a C programming problem. Reason about what kind of bugs students may make while coming up with solutions for the given problem.
        Next, come up with exactly 10 buggy implementations of the function {\color{OliveGreen} \{function\_name\}}, their corrected versions, and explanations for the bugs.
        Format it as a JSON object, where each object contains the following keys: `code', `fixed\_code', and `explanation':\newline
        \{\newline
        \null\quad"reasoning": "Reasoning about the bugs",\newline
        \null\quad"content": \newline
        \null\quad\quad[\{ \ "code": ...,\newline
        \null\quad\quad \ \ \ \ "fixed\_code": ...,\newline
        \null\quad\quad \ \ \ \ "explanation": ...\ \}]\newline
        \}\newline
        \vspace{-2mm}
        \newline
        Implement only this function with various bugs that students may make, incorporating the bugs you reasoned about. Each program should contain only one bug. Make them as diverse as possible. The bugs should not lead to the program not compiling or hanging.
        Do not add comments. The `stdio.h' library is already included and other libraries are not allowed.
        Do not forget to first reason about possible bugs.\newline
        \vspace{-2.25mm}
        \newline
        Problem Description:
        {\color{OliveGreen} \{problem\_specification\}}
        \\
        \hline
    \end{tabular}
    \vspace{-2.25mm}
     \caption{Prompt for asking LLMs to generate buggy codes.
     }
    \label{fig.prompt}
    \vspace{-2mm}
\end{figure}

\looseness-1\textbf{Implementation details.} It is desirable to have a diverse set of debugging exercises for a student to work on. For this, we ask the LLM to generate $10$ tuples consisting of a buggy code, its fixed version, and corresponding explanation, as illustrated by the prompt shown in Figure~\ref{fig.prompt}. We use these tuples to create $10$ exercises and keep only those that pass the validation stage. We deliver to the student the first exercise that passes validation, while caching the rest for subsequent requests.  We use a temperature of $0.7$ to introduce variability in the generated exercises, ensuring a diverse range of generations that simulate a wide array of potential bugs.


\section{Evaluation Procedure}\label{sec.setup}
\looseness-1In this section, we describe our evaluation of BugSpotter which includes an expert assessment of generated exercises and a large-scale classroom deployment to compare exercises generated by BugSpotter with instructor-created exercises. The evaluation is done over $3$ different problems, shown in Figures~\ref{fig.illustration_p1.description},~\ref{fig.illustration_p2.description},~and~\ref{fig.illustration_p3.description}.

\begin{figure*}[t!]
\centering{
\scalebox{0.82}{
\setlength\tabcolsep{7pt}
\renewcommand{\arraystretch}{0.92}  
    \begin{tabular}{l | c | c | c | r | r r | rrr}
    \toprule 
    \multirow[c]{2}{*}[0mm]{{\backslashbox[32mm]{\textbf{LLM:Problem}}{\textbf{Metric}}}} &
    \multicolumn{1}{c|}{\textbf{SuccOf10}} &
    \multicolumn{1}{c|}{\textbf{DiverseCodes}} &
    \multicolumn{1}{c|}{\textbf{BugProbRelated}} &
    \multicolumn{1}{c|}{\textbf{NbBugs}} &
    \multicolumn{2}{c|}{\textbf{EditTokens}} &
    \multicolumn{3}{c}{\textbf{BugType}} \\
    &
    &
    &
    &
    &
    \multicolumn{1}{c}{\textbf{Average}} &
    \multicolumn{1}{c|}{\textbf{Median}} &
    \multicolumn{1}{c}{\textbf{Type 1}} &
    \multicolumn{1}{c}{\textbf{Type 2}} &
    \multicolumn{1}{c}{\textbf{Type 3}} \\
    \midrule
    \midrule
    \TechGPTFour{}:Problem 1 &
    $6.33$ {\color{gray} $(2.0)$} &
    $5.67$ {\color{gray} $(1.7)$} &
    $1.00$ {\color{gray} $(0.0)$} &
    $1.04$ {\color{gray} $(0.1)$} &
    $3.13$ {\color{gray} $(\phantom{0} 0.9)$} &
    $1.83$ {\color{gray} $(\phantom{0} 0.5)$} &
    $0.57$ {\color{gray} $(0.2)$} &
    $0.39$ {\color{gray} $(0.2)$} &
    $0.04$ {\color{gray} $(0.1)$} \\
    \TechGPTFour{}:Problem 2 &
    $4.00$ {\color{gray} $(1.4)$} &
    $3.67$ {\color{gray} $(1.5)$} &
    $1.00$ {\color{gray} $(0.0)$} &
    $1.00$ {\color{gray} $(0.0)$} &
    $12.39$ {\color{gray} $(\phantom{0} 3.1)$} &
    $9.67$ {\color{gray} $(\phantom{0} 5.2)$} &
    $1.00$ {\color{gray} $(0.0)$} &
    $0.00$ {\color{gray} $(0.0)$} &
    $0.00$ {\color{gray} $(0.0)$} \\
    \TechGPTFour{}:Problem 3 &
    $5.67$ {\color{gray} $(0.4)$} &
    $5.33$ {\color{gray} $(0.4)$} &
    $1.00$ {\color{gray} $(0.0)$} &
    $1.29$ {\color{gray} $(0.1)$} &
    $12.37$ {\color{gray} $(\phantom{0} 0.8)$} &
    $8.17$ {\color{gray} $(\phantom{0} 2.0)$} &
    $0.94$ {\color{gray} $(0.1)$} &
    $0.00$ {\color{gray} $(0.0)$} &
    $0.06$ {\color{gray} $(0.1)$} \\

    \midrule
    \TechGPTFour{}:All &
    $5.33$ {\color{gray} $(0.7)$} &
    $4.89$ {\color{gray} $(0.8)$} &
    $1.00$ {\color{gray} $(0.0)$} &
    $1.11$ {\color{gray} $(0.0)$} &
    $9.29$ {\color{gray} $(\phantom{0} 0.8)$} &
    $6.56$ {\color{gray} $(\phantom{0} 1.9)$} &
    $0.84$ {\color{gray} $(0.1)$} &
    $0.13$ {\color{gray} $(0.1)$} &
    $0.03$ {\color{gray} $(0.0)$} \\

    \midrule
    \midrule
    \TechChatGPT{}:Problem 1 &
    $6.67$ {\color{gray} $(0.4)$} &
    $5.67$ {\color{gray} $(0.4)$} &
    $1.00$ {\color{gray} $(0.0)$} &
    $1.06$ {\color{gray} $(0.1)$} &
    $4.23$ {\color{gray} $(\phantom{0} 2.2)$} &
    $3.00$ {\color{gray} $(\phantom{0} 1.2)$} &
    $0.64$ {\color{gray} $(0.1)$} &
    $0.36$ {\color{gray} $(0.1)$} &
    $0.00$ {\color{gray} $(0.0)$} \\

    \TechChatGPT{}:Problem 2 &
    $4.67$ {\color{gray} $(1.6)$} &
    $3.50$ {\color{gray} $(1.1)$} &
    $1.00$ {\color{gray} $(0.0)$} &
    $1.00$ {\color{gray} $(0.0)$} &
    $23.00$ {\color{gray} $(17.2)$} &
    $22.17$ {\color{gray} $(16.9)$} &
    $1.00$ {\color{gray} $(0.0)$} &
    $0.00$ {\color{gray} $(0.0)$} &
    $0.00$ {\color{gray} $(0.0)$} \\

    \TechChatGPT{}:Problem 3 &
    $7.00$ {\color{gray} $(3.1)$} &
    $4.83$ {\color{gray} $(2.1)$} &
    $0.93$ {\color{gray} $(0.1)$} &
    $2.36$ {\color{gray} $(0.3)$} &
    $46.23$ {\color{gray} $(\phantom{0} 3.9)$} &
    $44.17$ {\color{gray} $(\phantom{0} 5.1)$} &
    $0.00$ {\color{gray} $(0.0)$} &
    $0.64$ {\color{gray} $(0.2)$} &
    $0.36$ {\color{gray} $(0.2)$} \\

    \midrule
    \TechChatGPT{}:All &
    $6.11$ {\color{gray} $(1.7)$} &
    $4.67$ {\color{gray} $(1.1)$} &
    $0.98$ {\color{gray} $(0.0)$} &
    $1.47$ {\color{gray} $(0.1)$} &
    $24.49$ {\color{gray} $(\phantom{0} 5.6)$} &
    $23.11$ {\color{gray} $(\phantom{0} 4.9)$} &
    $0.55$ {\color{gray} $(0.0)$} &
    $0.33$ {\color{gray} $(0.1)$} &
    $0.12$ {\color{gray} $(0.1)$} \\
    \bottomrule
    \end{tabular}
    } 
    } 
     \vspace{-2mm}
     \caption{
     Results of the expert-based quality assessment. BugSpotter leverages LLMs from OpenAI's GPT family \cite{GPT-Family}. Evaluation was done across 3 different problems, over 3 independent runs, according to the rubric described in Section~\ref{sec.setup}.
     }
     \vspace{-1mm}
    \label{fig.rq1}
\end{figure*}

\subsection{Expert Evaluation for RQ1}
\looseness-1Our first research question, RQ1, involves expert-based assessment to evaluate the quality of exercises produced by BugSpotter. We assess the quality w.r.t. several attributes reflecting their suitability for being used in the classroom. For this, we created the following rubric grounded in literature 
\cite{DBLP:journals/corr/abs-2306-17156,Ma_2024}, though adapted to capture more features of the exercises generated with BugSpotter. \emph{SuccOf10} ($0$ to $10$) reports how many of the $10$ generated exercises successfully pass validation. \emph{DiverseCodes} ($0$ to $10$) reports how many of the generated exercises that pass the validation stage contain unique bugs. \emph{BugProbRelated} (binary) is $1$ when the buggy code \buggy{} is an attempt at solving problem \problem{}, and $0$ when it is not related to \problem{} (i.e., trying to solve a different problem); we compute this attribute only for the exercises passing validation. \emph{NbBugs} (positive number) reports the number of conceptually different bugs present in \buggy{}; we compute this attribute only for the exercises passing validation. \emph{EditTokens} (non-negative number) captures the edit distance between strings obtained by tokenizing \buggy{} and \fixed{} using the Pygments library \cite{pygments}; we compute this attribute only for the exercises passing validation. \emph{BugType} (a one-hot encoded vector) characterizes the kind of behavior \buggy{} exhibits when executed on problem \problem{}'s test suite; we compute this attribute only for the exercises passing validation. In particular, \emph{BugType} categorizes exercises into the following three types:
\setlength{\leftmargini}{1em}
\begin{itemize}
    \item \emph{Type 1} means that buggy code \buggy{} passes at least one test case from problem \problem{}'s test suite, and does not result in a run-time error or division by $0$ on any test case (see Figure~\ref{fig.illustration_p2}).
    \item \emph{Type 2} means that buggy code \buggy{} does not pass any of the test cases from problem \problem{}'s test suite, and does not result in a run-time error or division by $0$ on any test case (see Figure~\ref{fig.illustration_p1}).
    \item \emph{Type 3} means that running buggy code \buggy{} results in a run-time error or division by $0$ on a test case (see Figure~\ref{fig.illustration_p3}).
\end{itemize}

\looseness-1Metrics \emph{SuccOf10}, \emph{EditTokens}, and \emph{BugType} can be computed automatically using scripts with no manual annotations. For the rest of the attributes, two human experts (evaluators) independently rated the generated exercises -- these two evaluators are experts in programming, with one evaluator having experience tutoring introductory-level programming courses. We obtained a Cohen's kappa reliability value greater than $0.7$ for each rated attribute, indicating \emph{substantial agreement} between evaluators \cite{cohen1960coefficient}. All results are reported based on average across annotations of the two evaluators.

\subsection{Classroom Evaluation for RQ2 and RQ3}
\looseness-1Our next two research questions, RQ2 and RQ3, involve assessing students' success on exercises generated with BugSpotter. We aim to understand the alignment between expert-classified difficulty of debugging exercises and the students' performance on the same exercises. Furthermore, we analyze whether there are significant differences between exercises generated with BugSpotter and exercises designed by instructors. To enable this, we deployed a version of BugSpotter as part of a laboratory task in a large introductory C programming course involving $741$ students, taught at the University of Auckland. Students in this course typically have no prior programming experience. The lab was conducted towards the end of the course, and students were assigned $3$ debugging exercises based on the $3$ different problems we used throughout our study.

\looseness-1For the classroom deployment, we first generated exercises with the BugSpotter pipeline employing various LLMs from the GPT family \cite{GPT-Family}. Then, we pre-selected $5$ exercises (instead of using real-time generation) to control the generation quality and avoid impacting the students' learning experience -- the pre-selection was done to ensure all the exercises are of high-quality and diverse based on the rubric established above. For RQ3, an instructor created an additional $5$ exercises by modifying reference solution code to include realistic bugs. As part of the deployment, students were assigned randomly to one of these $10$ exercises ($5$ LLM-generated and $5$ instructed-created). To explore the impact of both difficulty and source of buggy code on student performance, we collected student responses (i.e., test cases) and automatically computed their success rates on debugging exercises for our evaluation.


\section{Results}\label{sec.results}

In this section, we discuss the results of the study centered around the research questions (RQs) introduced in Section~\ref{sec.introduction}.

\subsection{RQ1: Expert-assessed Quality and Diversity}
We present results in terms of quality and diversity of generated exercises. Figure~\ref{fig.rq1} shows the performance of BugSpotter when using \TechChatGPT{} \cite{ChatGPT} and \TechGPTFour{} \cite{GPT4o} as LLMs for generating buggy codes. We conduct $3$ independent runs to report averaged results as \emph{mean (stderr)} for all attributes based on annotations by two evaluators. Our results show that \TechChatGPT{} is comparable with \TechGPTFour{} in terms of creating a diverse set of debugging exercises that are well-matched to the problem. Furthermore, \TechChatGPT{}-generated exercises pass the validation step more often than those created by \TechGPTFour{}. Next, we see that \TechGPTFour{} predominantly creates exercises with buggy codes that pass at least one of the problem's test cases (i.e., Type 1), while \TechChatGPT{} has a higher tendency to create buggy codes that pass none or result in an error, bringing more diversity w.r.t. the types of produced bugs. In terms of diversity of exercises passing validation, both LLMs are comparable for creating sets of diverse buggy codes. Finally, we  notice that \TechChatGPT{} has a slightly higher number of conceptually different bugs per buggy program than \TechGPTFour{}. This leads to \TechChatGPT{}-produced buggy code requiring more edits to fix, signaling bugs that are more difficult. 

\looseness-1These results highlight that both LLMs are suitable for the debugging exercise generation pipeline. Surprisingly, \TechChatGPT{}, the cheaper alternative, can produce exercises comparable in quality to \TechGPTFour{}. This means that \TechChatGPT{} can be used to power BugSpotter as a scalable solution in terms of cost, without compromising quality.

\subsection{RQ2: Expert-assessed Difficulty and Student Performance on Debugging Exercises}

\looseness-1For each problem, we analyze the $5$ pre-selected exercises generated by LLMs as described in Section~\ref{sec.setup}. Difficulty is often a key factor in deciding whether exercises should be assigned to students \cite{DBLP:conf/icwl/LabajB14}, so it is important for BugSpotter to produce a range of difficulties from which students can choose. We study this diversity from both the perspective of an expert and empirically based on student performance. After ranking the $5$ exercises from least difficult to most difficult, we label the first $2$ as easy, the next $2$ as medium, and the last as hard. Student performance is measured through success rates on exercises.

\begin{figure}[t!]
\centering
	\includegraphics[width=0.86\linewidth]{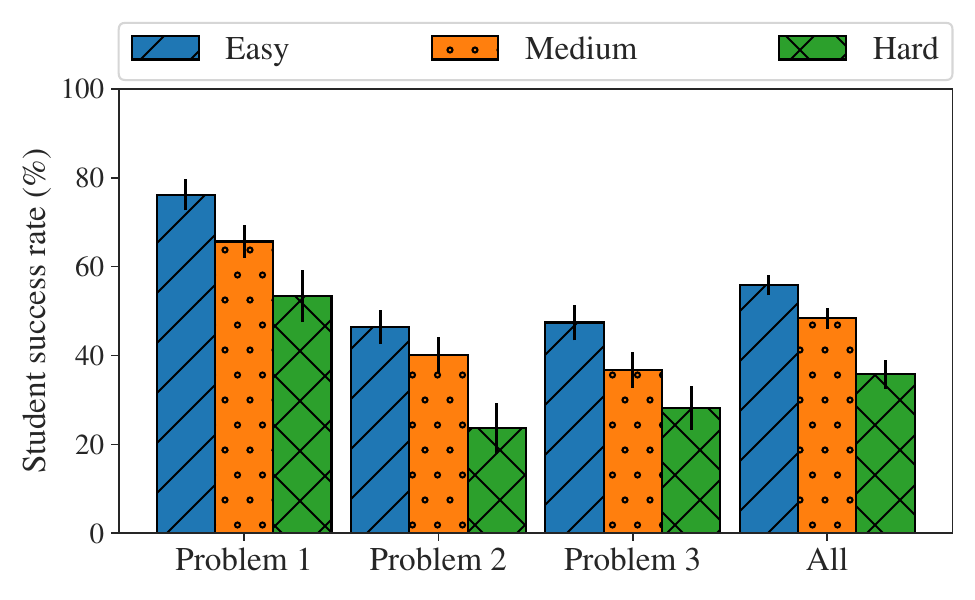}
    \vspace{-3.2mm}
     \caption{Student success rates w.r.t. expert-assessed difficulty of exercises. Aggregated per problem, success rates are 67.2\% for Problem 1, 40.5\% for Problem 2, and 39.2\% for Problem 3. 
     }
     \vspace{-2.5mm}
    \label{fig.rq2}
\end{figure}

Figure~\ref{fig.rq2} illustrates the correlation between success rates and expert-assessed difficulty. The results indicate that for all problems, exercises classified as easy have the highest success rates, followed by medium difficulty exercises, and then hard exercises. This pattern remains consistent when aggregating results across all problems. We believe that these results show a strong alignment between the expert assessment and student performance. More importantly, they show that LLMs can generate diverse sets of debugging exercises.

\subsection{RQ3: LLM-Generated vs. Instructor-Created Debugging Exercises}
\begin{figure}[t!]
\centering
	\includegraphics[width=0.86\linewidth]{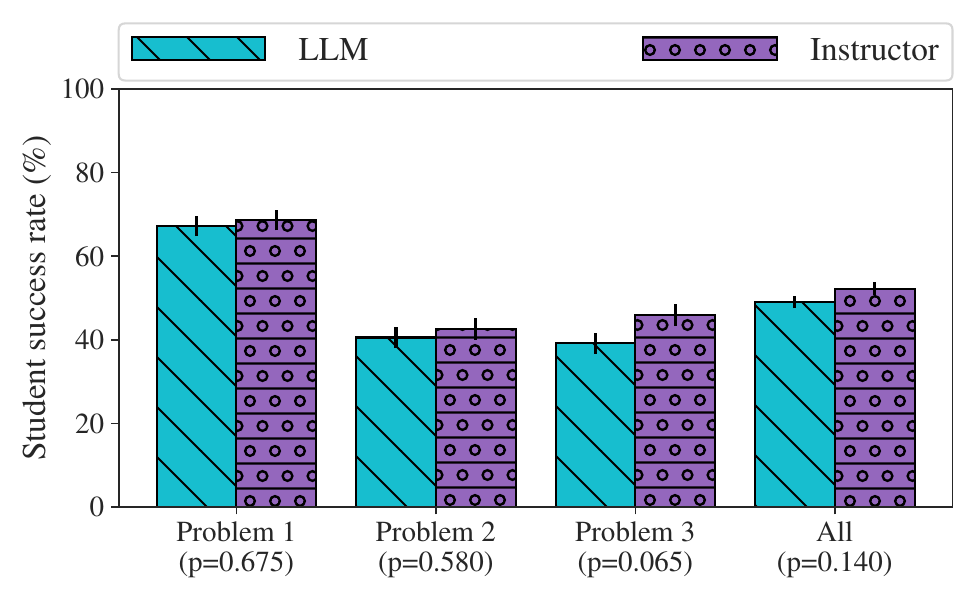}
     \vspace{-3.2mm}
     \caption{Comparison between student success rates on LLM-generated debugging exercises compared to student success rates on instructor-created exercises.
     }
     \vspace{-2.5mm}
    \label{fig.rq3}
\end{figure}

Next, we compare students' performance on the debugging exercises containing buggy codes generated by LLMs from the GPT family with those handcrafted by an instructor. We aim to explore whether there is any significant difference in students' performance in solving debugging exercises, depending on the source of the buggy code. Again, we analyze students' performance in terms of their success rate. Figure~\ref{fig.rq3} shows students' success rate for each problem and aggregated over all problems. The results show that the success rates for debugging exercises with codes created by the instructor are slightly higher, indicating that they are less difficult. To check whether this slight difference is significant, we compare students' performance for LLM-generated exercises with instructor-created exercises via the $\chi^{2}$ test \cite{cochran1952chi2}, using contingency tables with two rows (source) and two columns ($741$ data points per problem mapped to successful/unsuccessful). As shown in Figure~\ref{fig.rq3}, all p-values exceed $0.05$, with the lowest p-value being $0.065$, indicating no significant difference between the two sources.

We believe that these results strongly suggest that debugging exercises generated with the help of LLMs are comparable in difficulty to those handcrafted by instructors. This shows the potential of using LLMs to automate the creation of such exercises in class, reducing the workload on educators. The slight difference in success rate, while not significant, may indicate that buggy codes created by the instructor are more naturalistic, aligning more closely with the types of errors students encounter while coding on their own.

\subsection{Web Application}

\begin{figure*}[t!]
\centering
	\begin{minipage}{\linewidth}
    {\ \ \ \
        \begin{subfigure}{0.45\linewidth}
        {\centering
            \includegraphics[width=\linewidth]{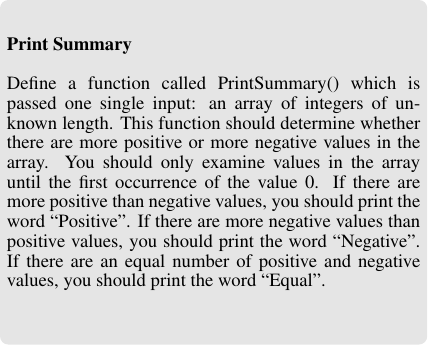}
            \vspace{-5.3mm}
            \caption{Problem specification}				
            \label{fig.illustration_p2.description}
        }
        \end{subfigure}
        \ \ \ \ \ \ \ \ \ \ \ \ \ \ \ \ \ \
        \begin{subfigure}{0.36\linewidth}
        {
            \setlength{\fboxsep}{2pt}\fbox{\includegraphics[width=\linewidth]{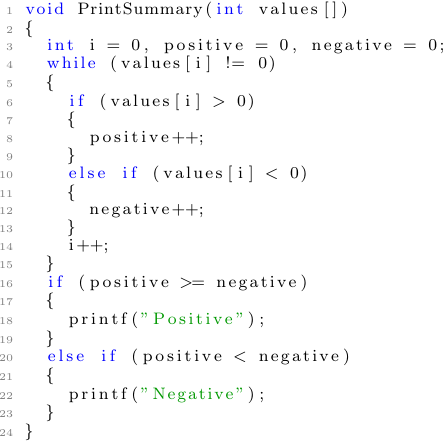}}
            \vspace{-4.5mm}
            \caption{Buggy code}				
            \label{fig.illustration_p2.code}
        }
        \end{subfigure}
    }
    \end{minipage}
    \begin{minipage}{\linewidth}
    {\
        \begin{subfigure}{0.95\linewidth}
        {\centering
            \vspace{0.1cm}
            \setlength{\fboxsep}{-2pt}\fbox{
            \includegraphics[width=\linewidth, trim=0mm 0.7mm 0mm 0.1mm]{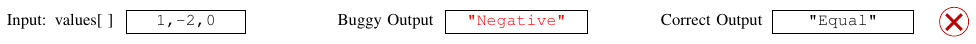}
            }
            \vspace{-5mm}
            \caption{Test case design}				
            \label{fig.illustration_p2.testcase}
        }
        \end{subfigure}
    }
    \end{minipage}
    \vspace{-4mm}
     \caption{\looseness-1Illustration of a debugging exercise for Problem 2. The exercise contains a Type 1 buggy code, as it passes some of the test cases in the problem's test suite. In this example, the student's attempt is wrong (cross \img{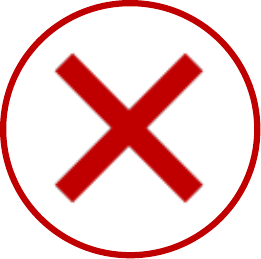}), as the provided buggy output does not match the actual output of the buggy code when run on the input, thus not meeting criterion (2) w.r.t. solving the exercise.
     }
     \vspace{-1.5mm}
    \label{fig.illustration_p2}
\end{figure*}

\begin{figure*}[t!]
\centering
    \begin{minipage}{\linewidth}
    {\
        \begin{subfigure}{0.5\linewidth}
        {\centering
            \includegraphics[width=0.95\linewidth]{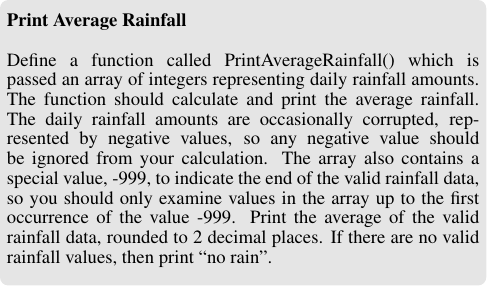}
            \vspace{-1mm}
            \caption{Problem specification}				
            \label{fig.illustration_p3.description}
        }
        \end{subfigure}
        \ \ \ \ \
        \begin{subfigure}{0.44\linewidth}
        {
            \setlength{\fboxsep}{2pt}\fbox{\includegraphics[width=0.95\linewidth]{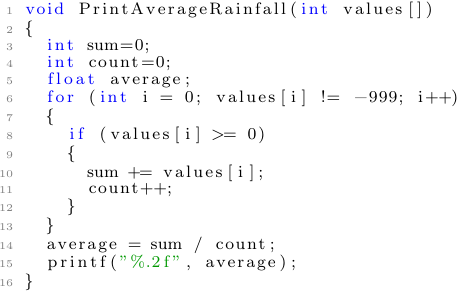}}
            \vspace{-1mm}
            \caption{Buggy code}				
            \label{fig.illustration_p3.code}
        }
        \end{subfigure}
    }
    \end{minipage}
    \begin{minipage}{\linewidth}
    {\ \
        \begin{subfigure}{0.95\linewidth}
        {\centering
            \vspace{0.1cm}
            \setlength{\fboxsep}{-2pt}\fbox{
            \includegraphics[width=\linewidth, trim=0mm 0.7mm 0mm 0.1mm]{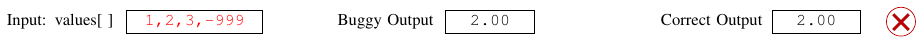}
            }
            \vspace{-5mm}
            \caption{Test case design}				
            \label{fig.illustration_p3.testcase}
        }
        \end{subfigure}
    }
    \end{minipage}
    \vspace{-4mm}
     \caption{ 
     Illustration of a debugging exercise for Problem 3. The exercise contains a Type 3 buggy code, as it does not check before dividing with a variable that may potentially be equal to 0. In this example, the student's attempt is wrong (cross \img{reject}), as the provided test case is not a failing test case, thus not meeting criterion (1) w.r.t. solving the exercise.
     }
     \vspace{-2.5mm}
    \label{fig.illustration_p3}
\end{figure*}

\looseness-1A free web application showing the capabilities of BugSpotter can be found at \url{https://bugspotter.netlify.app/}. This web application provides a demo similar to the format of the illustrative examples shown in Figures~\ref{fig.illustration_p1},~\ref{fig.illustration_p2}~and~\ref{fig.illustration_p3}. After providing a test case, a student can submit their answer for automatic validation. They can also request additional debugging exercises with the press of a button -- after a short wait, BugSpotter delivers a new exercise. This demo application supports Python debugging exercises and we plan to add support for C debugging exercises in the future.

\subsection{Limitations}
\looseness-1Next, we discuss a few limitations of our study. 
Firstly, our classroom investigation was done using pre-selected exercises (instead of using real-time generation) to control the generation quality for the laboratory task. For future studies, it would be more informative to analyze students' performance on buggy exercises generated with LLMs in real-time. 
Secondly, we did not correlate students' success rates for debugging exercises with their long-term course performance. This would be a good indicator of the exercises' suitability for being included in classroom assignments.
Thirdly, our study did not capture how BugSpotter could help educators in terms of reducing their workload in creating debugging exercises. This would be needed to fully understand the practical utility of BugSpotter.


\section{Concluding Discussion}\label{sec.conclusion}

\looseness-1We developed BugSpotter, a tool for debugging practice with LLM-generated buggy code for problems that require implementing a single function. We tested the generated exercises' quality via both expert-based and classroom evaluations. As a first finding, our results show that with minor differences, both \TechGPTFour{} and \TechChatGPT{} produce diverse sets of buggy codes that can be leveraged by BugSpotter's debugging exercise generation pipeline. Our second finding is that there is a high alignment between expert-assessed difficulty and students' performance on generated exercises. This underlines that LLMs are suitable for generating diverse sets of exercises with varying difficulty. Our third finding obtained through further investigation of the classroom setting shows that exercises generated by LLMs are comparable in difficulty to those created by instructors. Statistical analysis shows no significant difference between the difficulty of LLM-generated and instructor-created exercises. We believe this signals great potential for using LLMs to automate the creation of debugging exercises. 

\looseness-1There are several exciting directions for future work. Firstly, it would be useful to extend BugSpotter to generate debugging exercises for problem specifications that involve multiple functions and complex programming concepts, e.g., involving classes or I/O files. Another direction would be to fine-tune LLMs to produce bugs more similar to those made by students -- such debugging exercises may better prepare students to tackle their own likely bugs. A broader direction is to add more types of debugging and test-writing exercises. Students could write several test cases to capture multiple bugs, edit code to fix them, or even submit their own problems and test suites, which tools like BugSpotter can use to generate new exercises.


\begin{acks}
Funded/Cofunded by the European Union (ERC, TOPS, 101039090). Views and opinions expressed are however those of the author(s) only and do not necessarily reflect those of the European Union or the European Research Council. Neither the European Union nor the granting authority can be held responsible for them.
\end{acks}

\bibliographystyle{ACM-Reference-Format}
\balance
\bibliography{main}


\begin{thebibliography}{41}


\ifx \showCODEN    \undefined \def \showCODEN     #1{\unskip}     \fi
\ifx \showDOI      \undefined \def \showDOI       #1{#1}\fi
\ifx \showISBNx    \undefined \def \showISBNx     #1{\unskip}     \fi
\ifx \showISBNxiii \undefined \def \showISBNxiii  #1{\unskip}     \fi
\ifx \showISSN     \undefined \def \showISSN      #1{\unskip}     \fi
\ifx \showLCCN     \undefined \def \showLCCN      #1{\unskip}     \fi
\ifx \shownote     \undefined \def \shownote      #1{#1}          \fi
\ifx \showarticletitle \undefined \def \showarticletitle #1{#1}   \fi
\ifx \showURL      \undefined \def \showURL       {\relax}        \fi
\providecommand\bibfield[2]{#2}
\providecommand\bibinfo[2]{#2}
\providecommand\natexlab[1]{#1}
\providecommand\showeprint[2][]{arXiv:#2}

\bibitem[Ahmed et~al\mbox{.}(2020)]%
        {DBLP:conf/nips/AhmedCEFGRS20}
\bibfield{author}{\bibinfo{person}{Umair~Z. Ahmed}, \bibinfo{person}{Maria Christakis}, \bibinfo{person}{Aleksandr Efremov}, \bibinfo{person}{Nigel Fernandez}, \bibinfo{person}{Ahana Ghosh}, \bibinfo{person}{Abhik Roychoudhury}, {and} \bibinfo{person}{Adish Singla}.} \bibinfo{year}{2020}\natexlab{}.
\newblock \showarticletitle{{S}ynthesizing {T}asks for {B}lock-based {P}rogramming}. In \bibinfo{booktitle}{\emph{Proceedings of the Annual Conference on Neural Information Processing Systems ({NeurIPS})}}.
\newblock


\bibitem[Brandl et~al\mbox{.}(2006)]%
        {pygments}
\bibfield{author}{\bibinfo{person}{Georg Brandl}, \bibinfo{person}{Matth{\"a}us Chajdas}, {and} \bibinfo{person}{Jean Abou-Samra}.} \bibinfo{year}{2006}\natexlab{}.
\newblock \bibinfo{title}{Pygments}.
\newblock \bibinfo{howpublished}{\url{https://pygments.org/}}.
\newblock


\bibitem[Caraco et~al\mbox{.}(2024)]%
        {DBLP:conf/sigcse/CaracoLVF24}
\bibfield{author}{\bibinfo{person}{Serena Caraco}, \bibinfo{person}{Nelson Lojo}, \bibinfo{person}{Michael Verdicchio}, {and} \bibinfo{person}{Armando Fox}.} \bibinfo{year}{2024}\natexlab{}.
\newblock \showarticletitle{{G}enerating {M}ulti-{P}art {A}utogradable {F}aded {P}arsons {P}roblems {F}rom {C}ode-{W}riting {E}xercises}. In \bibinfo{booktitle}{\emph{Proceedings of the Technical Symposium on Computer Science Education ({SIGCSE})}}.
\newblock


\bibitem[Chen et~al\mbox{.}(2013)]%
        {DBLP:conf/latice/ChenWL13}
\bibfield{author}{\bibinfo{person}{Mei{-}Wen Chen}, \bibinfo{person}{Cheng{-}Chih Wu}, {and} \bibinfo{person}{Yu{-}Tzu Lin}.} \bibinfo{year}{2013}\natexlab{}.
\newblock \showarticletitle{{N}ovices' {D}ebugging {B}ehaviors in {VB} {P}rogramming}. In \bibinfo{booktitle}{\emph{Proceedings of the Learning and Teaching in Computing and Engineering ({LaTiCE})}}.
\newblock


\bibitem[Cochran(1952)]%
        {cochran1952chi2}
\bibfield{author}{\bibinfo{person}{William~G Cochran}.} \bibinfo{year}{1952}\natexlab{}.
\newblock \showarticletitle{The $\chi$2 {T}est of {G}oodness of {F}it}.
\newblock \bibinfo{journal}{\emph{The Annals of Mathematical Statistics}} (\bibinfo{year}{1952}).
\newblock


\bibitem[Cohen(1960)]%
        {cohen1960coefficient}
\bibfield{author}{\bibinfo{person}{Jacob Cohen}.} \bibinfo{year}{1960}\natexlab{}.
\newblock \showarticletitle{{A} {C}oefficient of {A}greement for {N}ominal {S}cales}.
\newblock \bibinfo{journal}{\emph{Educational and Psychological Measurement}} (\bibinfo{year}{1960}).
\newblock


\bibitem[Denny et~al\mbox{.}(2024a)]%
        {DBLP:journals/corr/abs-2402-01580}
\bibfield{author}{\bibinfo{person}{Paul Denny}, \bibinfo{person}{Sumit Gulwani}, \bibinfo{person}{Neil~T. Heffernan}, \bibinfo{person}{Tanja K{\"{a}}ser}, \bibinfo{person}{Steven Moore}, \bibinfo{person}{Anna~N. Rafferty}, {and} \bibinfo{person}{Adish Singla}.} \bibinfo{year}{2024}\natexlab{a}.
\newblock \showarticletitle{{G}enerative {AI} for {E}ducation {(GAIED):} {A}dvances, {O}pportunities, and {C}hallenges}.
\newblock \bibinfo{journal}{\emph{CoRR}}  \bibinfo{volume}{abs/2402.01580} (\bibinfo{year}{2024}).
\newblock


\bibitem[Denny et~al\mbox{.}(2024b)]%
        {DBLP:conf/sigcse/00010PLABR24}
\bibfield{author}{\bibinfo{person}{Paul Denny}, \bibinfo{person}{Juho Leinonen}, \bibinfo{person}{James Prather}, \bibinfo{person}{Andrew Luxton{-}Reilly}, \bibinfo{person}{Thezyrie Amarouche}, \bibinfo{person}{Brett~A. Becker}, {and} \bibinfo{person}{Brent~N. Reeves}.} \bibinfo{year}{2024}\natexlab{b}.
\newblock \showarticletitle{{P}rompt {P}roblems: {A} {N}ew {P}rogramming {E}xercise for the {G}enerative {AI} {E}ra}. In \bibinfo{booktitle}{\emph{Proceedings of the Technical Symposium on Computer Science Education ({SIGCSE})}}.
\newblock


\bibitem[Denny et~al\mbox{.}(2011)]%
        {DBLP:conf/sigcse/DennyLTH11}
\bibfield{author}{\bibinfo{person}{Paul Denny}, \bibinfo{person}{Andrew Luxton{-}Reilly}, \bibinfo{person}{Ewan~D. Tempero}, {and} \bibinfo{person}{Jacob Hendrickx}.} \bibinfo{year}{2011}\natexlab{}.
\newblock \showarticletitle{{CodeWrite}: {S}upporting {S}tudent-{D}riven {P}ractice of {J}ava}. In \bibinfo{booktitle}{\emph{Proceedings of the Technical Symposium on Computer Science Education ({SIGCSE})}}.
\newblock


\bibitem[Denny et~al\mbox{.}(2019)]%
        {denny2019closer}
\bibfield{author}{\bibinfo{person}{Paul Denny}, \bibinfo{person}{James Prather}, \bibinfo{person}{Brett~A. Becker}, \bibinfo{person}{Zachary Albrecht}, \bibinfo{person}{Dastyni Loksa}, {and} \bibinfo{person}{Raymond Pettit}.} \bibinfo{year}{2019}\natexlab{}.
\newblock \showarticletitle{{A} {C}loser {L}ook at {M}etacognitive {S}caffolding: {S}olving {T}est {C}ases {B}efore {P}rogramming}. In \bibinfo{booktitle}{\emph{Koli Calling International Conference on Computing Education Research ({Koli Calling})}}.
\newblock


\bibitem[Denny et~al\mbox{.}(2024c)]%
        {denny2024cacm}
\bibfield{author}{\bibinfo{person}{Paul Denny}, \bibinfo{person}{James Prather}, \bibinfo{person}{Brett~A. Becker}, \bibinfo{person}{James Finnie{-}Ansley}, \bibinfo{person}{Arto Hellas}, \bibinfo{person}{Juho Leinonen}, \bibinfo{person}{Andrew Luxton{-}Reilly}, \bibinfo{person}{Brent~N. Reeves}, \bibinfo{person}{Eddie~Antonio Santos}, {and} \bibinfo{person}{Sami Sarsa}.} \bibinfo{year}{2024}\natexlab{c}.
\newblock \showarticletitle{{C}omputing {E}ducation in the {E}ra of {G}enerative {AI}}.
\newblock \bibinfo{journal}{\emph{Commun. ACM}} (\bibinfo{year}{2024}).
\newblock


\bibitem[Fitzgerald et~al\mbox{.}(2008)]%
        {fitzgerald2008debugging}
\bibfield{author}{\bibinfo{person}{Sue Fitzgerald}, \bibinfo{person}{Gary Lewandowski}, \bibinfo{person}{Ren{\'{e}}e McCauley}, \bibinfo{person}{Laurie Murphy}, \bibinfo{person}{Beth Simon}, \bibinfo{person}{Lynda Thomas}, {and} \bibinfo{person}{Carol Zander}.} \bibinfo{year}{2008}\natexlab{}.
\newblock \showarticletitle{{D}ebugging: {F}inding, {F}ixing and {F}lailing, a {M}ulti-institutional {S}tudy of {N}ovice {D}ebuggers}.
\newblock \bibinfo{journal}{\emph{Computer Science Education}}  \bibinfo{volume}{18} (\bibinfo{year}{2008}).
\newblock


\bibitem[Ghosh et~al\mbox{.}(2022)]%
        {DBLP:conf/aied/GhoshTDS22}
\bibfield{author}{\bibinfo{person}{Ahana Ghosh}, \bibinfo{person}{Sebastian Tschiatschek}, \bibinfo{person}{Sam Devlin}, {and} \bibinfo{person}{Adish Singla}.} \bibinfo{year}{2022}\natexlab{}.
\newblock \showarticletitle{{A}daptive {S}caffolding in {B}lock-{B}ased {P}rogramming via {S}ynthesizing {N}ew {T}asks as {P}op {Q}uizzes}. In \bibinfo{booktitle}{\emph{Proceeding of the International Conference on Artificial Intelligence in Education {AIED}}}.
\newblock


\bibitem[Gutierrez et~al\mbox{.}(2024)]%
        {DBLP:conf/sigcse/Gutierrez0L24}
\bibfield{author}{\bibinfo{person}{Andre Del~Carpio Gutierrez}, \bibinfo{person}{Paul Denny}, {and} \bibinfo{person}{Andrew Luxton{-}Reilly}.} \bibinfo{year}{2024}\natexlab{}.
\newblock \showarticletitle{{E}valuating {A}utomatically {G}enerated {C}ontextualised {P}rogramming {E}xercises}. In \bibinfo{booktitle}{\emph{Proceedings of the Technical Symposium on Computer Science Education ({SIGCSE})}}.
\newblock


\bibitem[Jordan et~al\mbox{.}(2024)]%
        {DBLP:conf/sigcse/JordanLR24}
\bibfield{author}{\bibinfo{person}{Mollie Jordan}, \bibinfo{person}{Kevin Ly}, {and} \bibinfo{person}{Adalbert Gerald~Soosai Raj}.} \bibinfo{year}{2024}\natexlab{}.
\newblock \showarticletitle{{N}eed a {P}rogramming {E}xercise {G}enerated in {Y}our {N}ative {L}anguage? {ChatGPT}'s {G}ot {Y}our {B}ack: {A}utomatic {G}eneration of {N}on-{E}nglish {P}rogramming {E}xercises {U}sing {OpenAI} {GPT-3.5}}. In \bibinfo{booktitle}{\emph{Proceedings of the Technical Symposium on Computer Science Education ({SIGCSE})}}.
\newblock


\bibitem[Kafai et~al\mbox{.}(2019)]%
        {DBLP:conf/sigcse/KafaiDFLC19}
\bibfield{author}{\bibinfo{person}{Yasmin~B. Kafai}, \bibinfo{person}{David DeLiema}, \bibinfo{person}{Deborah~A. Fields}, \bibinfo{person}{Gary Lewandowski}, {and} \bibinfo{person}{Colleen~M. Lewis}.} \bibinfo{year}{2019}\natexlab{}.
\newblock \showarticletitle{{R}ethinking {D}ebugging as {P}roductive {F}ailure for {CS} {E}ducation}. In \bibinfo{booktitle}{\emph{Proceedings of the Technical Symposium on Computer Science Education ({SIGCSE})}}.
\newblock


\bibitem[Ko et~al\mbox{.}(2019)]%
        {DBLP:conf/sigcse/KoLHKKQAP19}
\bibfield{author}{\bibinfo{person}{Amy~J. Ko}, \bibinfo{person}{Thomas~D. LaToza}, \bibinfo{person}{Stephen Hull}, \bibinfo{person}{Ellen~A. Ko}, \bibinfo{person}{William Kwok}, \bibinfo{person}{Jane Quichocho}, \bibinfo{person}{Harshitha Akkaraju}, {and} \bibinfo{person}{Rishin Pandit}.} \bibinfo{year}{2019}\natexlab{}.
\newblock \showarticletitle{{T}eaching {E}xplicit {P}rogramming {S}trategies to {A}dolescents}. In \bibinfo{booktitle}{\emph{Proceedings of the Technical Symposium on Computer Science Education ({SIGCSE})}}.
\newblock


\bibitem[Kotalwar et~al\mbox{.}(2024)]%
        {kotalwar2024hintsinbrowser}
\bibfield{author}{\bibinfo{person}{Nachiket Kotalwar}, \bibinfo{person}{Alkis Gotovos}, {and} \bibinfo{person}{Adish Singla}.} \bibinfo{year}{2024}\natexlab{}.
\newblock \showarticletitle{{Hints-In-Browser}: {B}enchmarking {L}anguage Models for {P}rogramming {F}eedback {G}eneration}. In \bibinfo{booktitle}{\emph{Proceedings of the Annual Conference on Neural Information Processing Systems ({NeurIPS}) Track on Datasets and Benchmarks}}.
\newblock


\bibitem[Labaj and Bielikov{\'{a}}(2014)]%
        {DBLP:conf/icwl/LabajB14}
\bibfield{author}{\bibinfo{person}{Martin Labaj} {and} \bibinfo{person}{M{\'{a}}ria Bielikov{\'{a}}}.} \bibinfo{year}{2014}\natexlab{}.
\newblock \showarticletitle{Utilization of Exercise Difficulty Rating by Students for Recommendation}. In \bibinfo{booktitle}{\emph{Proceedings of the International Conference on Web-Based Learning {(ICWL)}}}.
\newblock


\bibitem[Leinonen et~al\mbox{.}(2023)]%
        {DBLP:conf/sigcse/0001HSRDPB23}
\bibfield{author}{\bibinfo{person}{Juho Leinonen}, \bibinfo{person}{Arto Hellas}, \bibinfo{person}{Sami Sarsa}, \bibinfo{person}{Brent~N. Reeves}, \bibinfo{person}{Paul Denny}, \bibinfo{person}{James Prather}, {and} \bibinfo{person}{Brett~A. Becker}.} \bibinfo{year}{2023}\natexlab{}.
\newblock \showarticletitle{{U}sing {L}arge {L}anguage {M}odels to {E}nhance {P}rogramming {E}rror {M}essages}. In \bibinfo{booktitle}{\emph{Proceedings of the Technical Symposium on Computer Science Education ({SIGCSE})}}.
\newblock


\bibitem[Li et~al\mbox{.}(2019)]%
        {li2019towards}
\bibfield{author}{\bibinfo{person}{Chen Li}, \bibinfo{person}{Emily Chan}, \bibinfo{person}{Paul Denny}, \bibinfo{person}{Andrew Luxton{-}Reilly}, {and} \bibinfo{person}{Ewan~D. Tempero}.} \bibinfo{year}{2019}\natexlab{}.
\newblock \showarticletitle{{T}owards a {F}ramework for {T}eaching {D}ebugging}. In \bibinfo{booktitle}{\emph{Proceedings of the Australasian Computing Education Conference {(ACE)}}}.
\newblock


\bibitem[Lojo and Fox(2022)]%
        {DBLP:conf/iticse/LojoF22}
\bibfield{author}{\bibinfo{person}{Nelson Lojo} {and} \bibinfo{person}{Armando Fox}.} \bibinfo{year}{2022}\natexlab{}.
\newblock \showarticletitle{{T}eaching {T}est-{W}riting {A}s a {V}ariably-{S}caffolded {P}rogramming {P}attern}. In \bibinfo{booktitle}{\emph{Proceedings of the Conference on Innovation and Technology in Computer Science Education ({ItiCSE})}}.
\newblock


\bibitem[Ma et~al\mbox{.}(2024)]%
        {Ma_2024}
\bibfield{author}{\bibinfo{person}{Qianou Ma}, \bibinfo{person}{Hua Shen}, \bibinfo{person}{Kenneth Koedinger}, {and} \bibinfo{person}{Sherry~Tongshuang Wu}.} \bibinfo{year}{2024}\natexlab{}.
\newblock \showarticletitle{{H}ow to {T}each {P}rogramming in the {AI} {E}ra? {U}sing {LLMs} as a {T}eachable {A}gent for {D}ebugging}. In \bibinfo{booktitle}{\emph{Proceeding of the International Conference on Artificial Intelligence in Education {(AIED)}}}.
\newblock


\bibitem[MacNeil et~al\mbox{.}(2024)]%
        {DBLP:conf/ace/MacNeil0T0BHSK24}
\bibfield{author}{\bibinfo{person}{Stephen MacNeil}, \bibinfo{person}{Paul Denny}, \bibinfo{person}{Andrew Tran}, \bibinfo{person}{Juho Leinonen}, \bibinfo{person}{Seth Bernstein}, \bibinfo{person}{Arto Hellas}, \bibinfo{person}{Sami Sarsa}, {and} \bibinfo{person}{Joanne Kim}.} \bibinfo{year}{2024}\natexlab{}.
\newblock \showarticletitle{{D}ecoding {L}ogic {E}rrors: {A} {C}omparative {S}tudy on {B}ug {D}etection by {S}tudents and {L}arge {L}anguage {M}odels}. In \bibinfo{booktitle}{\emph{Proceedings of the Australasian Computing Education Conference {(ACE)}}}.
\newblock


\bibitem[McCauley et~al\mbox{.}(2008)]%
        {mccauley2008debugging}
\bibfield{author}{\bibinfo{person}{Ren{\'{e}}e McCauley}, \bibinfo{person}{Sue Fitzgerald}, \bibinfo{person}{Gary Lewandowski}, \bibinfo{person}{Laurie Murphy}, \bibinfo{person}{Beth Simon}, \bibinfo{person}{Lynda Thomas}, {and} \bibinfo{person}{Carol Zander}.} \bibinfo{year}{2008}\natexlab{}.
\newblock \showarticletitle{{D}ebugging: {A} {R}eview of the {L}iterature from an {E}ducational {P}erspective}.
\newblock \bibinfo{journal}{\emph{Computer Science Education}} (\bibinfo{year}{2008}).
\newblock


\bibitem[Molina et~al\mbox{.}(2024)]%
        {molina2024leveragingllmtutoringsystems}
\bibfield{author}{\bibinfo{person}{Ismael~Villegas Molina}, \bibinfo{person}{Audria Montalvo}, \bibinfo{person}{Benjamin Ochoa}, \bibinfo{person}{Paul Denny}, {and} \bibinfo{person}{Leo Porter}.} \bibinfo{year}{2024}\natexlab{}.
\newblock \bibinfo{title}{{L}everaging {LLM} {T}utoring {S}ystems for {N}on-{N}ative {E}nglish {S}peakers in {I}ntroductory {CS} {C}ourses}.
\newblock
\newblock


\bibitem[Murphy et~al\mbox{.}(2008)]%
        {DBLP:conf/sigcse/MurphyLMSTZ08}
\bibfield{author}{\bibinfo{person}{Laurie Murphy}, \bibinfo{person}{Gary Lewandowski}, \bibinfo{person}{Ren{\'{e}}e McCauley}, \bibinfo{person}{Beth Simon}, \bibinfo{person}{Lynda Thomas}, {and} \bibinfo{person}{Carol Zander}.} \bibinfo{year}{2008}\natexlab{}.
\newblock \showarticletitle{{D}ebugging: the {G}ood, the {B}ad, and the {Q}uirky -- {A} {Q}ualitative {A}nalysis of {N}ovices' {S}trategies}. In \bibinfo{booktitle}{\emph{Proceedings of the Technical Symposium on Computer Science Education ({SIGCSE})}}.
\newblock


\bibitem[Nguyen et~al\mbox{.}(2024)]%
        {DBLP:journals/corr/abs-2310-10690}
\bibfield{author}{\bibinfo{person}{Manh~Hung Nguyen}, \bibinfo{person}{Sebastian Tschiatschek}, {and} \bibinfo{person}{Adish Singla}.} \bibinfo{year}{2024}\natexlab{}.
\newblock \showarticletitle{{L}arge {L}anguage {M}odels for {I}n-{C}ontext {S}tudent {M}odeling: {S}ynthesizing {S}tudent's {B}ehavior in {V}isual {P}rogramming from {O}ne-{S}hot {O}bservation}. In \bibinfo{booktitle}{\emph{Proceedings of the International Conference on Educational Data Mining ({EDM})}}.
\newblock


\bibitem[OpenAI(2023)]%
        {ChatGPT}
\bibfield{author}{\bibinfo{person}{OpenAI}.} \bibinfo{year}{2023}\natexlab{}.
\newblock \bibinfo{title}{Chat{GPT}}.
\newblock \bibinfo{howpublished}{\url{https://openai.com/blog/chatgpt}}.
\newblock


\bibitem[OpenAI(2024a)]%
        {GPT4o}
\bibfield{author}{\bibinfo{person}{OpenAI}.} \bibinfo{year}{2024}\natexlab{a}.
\newblock \bibinfo{title}{{H}ello {GPT-4o}}.
\newblock \bibinfo{howpublished}{\url{https://openai.com/index/hello-gpt-4o/}}.
\newblock


\bibitem[OpenAI(2024b)]%
        {GPT-Family}
\bibfield{author}{\bibinfo{person}{OpenAI}.} \bibinfo{year}{2024}\natexlab{b}.
\newblock \bibinfo{title}{{O}pen{AI} {P}latform {M}odels}.
\newblock \bibinfo{howpublished}{\url{https://platform.openai.com/docs/models}}.
\newblock


\bibitem[P{\u{a}}durean and Singla(2024)]%
        {padurean24benchmarking}
\bibfield{author}{\bibinfo{person}{Victor-Alexandru P{\u{a}}durean} {and} \bibinfo{person}{Adish Singla}.} \bibinfo{year}{2024}\natexlab{}.
\newblock \showarticletitle{{B}enchmarking {G}enerative {M}odels on {C}omputational {T}hinking {T}ests in {E}lementary {V}isual {P}rogramming}. In \bibinfo{booktitle}{\emph{Proceedings of the Annual Conference on Neural Information Processing Systems ({NeurIPS}) Track on Datasets and Benchmarks}}.
\newblock


\bibitem[P{\u{a}}durean et~al\mbox{.}(2024)]%
        {padurean2024neural}
\bibfield{author}{\bibinfo{person}{Victor-Alexandru P{\u{a}}durean}, \bibinfo{person}{Georgios Tzannetos}, {and} \bibinfo{person}{Adish Singla}.} \bibinfo{year}{2024}\natexlab{}.
\newblock \showarticletitle{{N}eural {T}ask {S}ynthesis for {V}isual {P}rogramming}.
\newblock \bibinfo{journal}{\emph{Transactions on Machine Learning Research ({TMLR})}} (\bibinfo{year}{2024}).
\newblock


\bibitem[Phung et~al\mbox{.}(2023a)]%
        {DBLP:conf/edm/PhungCGKMSS23}
\bibfield{author}{\bibinfo{person}{Tung Phung}, \bibinfo{person}{Jos{\'{e}} Cambronero}, \bibinfo{person}{Sumit Gulwani}, \bibinfo{person}{Tobias Kohn}, \bibinfo{person}{Rupak Majumdar}, \bibinfo{person}{Adish Singla}, {and} \bibinfo{person}{Gustavo Soares}.} \bibinfo{year}{2023}\natexlab{a}.
\newblock \showarticletitle{{G}enerating {H}igh-{P}recision {F}eedback for {P}rogramming {S}yntax {E}rrors using {L}arge {L}anguage {M}odels}. In \bibinfo{booktitle}{\emph{Proceedings of the International Conference on Educational Data Mining {(EDM)}}}.
\newblock


\bibitem[Phung et~al\mbox{.}(2023b)]%
        {DBLP:journals/corr/abs-2306-17156}
\bibfield{author}{\bibinfo{person}{Tung Phung}, \bibinfo{person}{Victor{-}Alexandru Padurean}, \bibinfo{person}{Jos{\'{e}} Cambronero}, \bibinfo{person}{Sumit Gulwani}, \bibinfo{person}{Tobias Kohn}, \bibinfo{person}{Rupak Majumdar}, \bibinfo{person}{Adish Singla}, {and} \bibinfo{person}{Gustavo Soares}.} \bibinfo{year}{2023}\natexlab{b}.
\newblock \showarticletitle{{G}enerative {AI} for {P}rogramming {E}ducation: {B}enchmarking {C}hatGPT, {G}PT-4, and {H}uman {T}utors}. In \bibinfo{booktitle}{\emph{Proceedings of the Conference on International Computing Education Research ({ICER}) - Volume 2}}.
\newblock


\bibitem[Phung et~al\mbox{.}(2024)]%
        {DBLP:conf/lak/PhungPS0CGSS24}
\bibfield{author}{\bibinfo{person}{Tung Phung}, \bibinfo{person}{Victor{-}Alexandru Padurean}, \bibinfo{person}{Anjali Singh}, \bibinfo{person}{Christopher Brooks}, \bibinfo{person}{Jos{\'{e}} Cambronero}, \bibinfo{person}{Sumit Gulwani}, \bibinfo{person}{Adish Singla}, {and} \bibinfo{person}{Gustavo Soares}.} \bibinfo{year}{2024}\natexlab{}.
\newblock \showarticletitle{{A}utomating {H}uman {T}utor-{S}tyle {P}rogramming {F}eedback: {L}everaging {GPT-4} {T}utor {M}odel for {H}int {G}eneration and {GPT-3.5} {S}tudent {M}odel for {H}int {V}alidation}. In \bibinfo{booktitle}{\emph{Proceedings of the Learning Analytics and Knowledge Conference ({LAK})}}.
\newblock


\bibitem[Sanders et~al\mbox{.}(2013)]%
        {DBLP:conf/iticse/SandersACEGJLMP13}
\bibfield{author}{\bibinfo{person}{Kate Sanders} {et~al\mbox{.}}} \bibinfo{year}{2013}\natexlab{}.
\newblock \showarticletitle{{T}he {C}anterbury {QuestionBank}: {B}uilding a {R}epository of {M}ultiple-{C}hoice {CS1} and {CS2} {Q}uestions}. In \bibinfo{booktitle}{\emph{Proceedings of the Working Group Reports of the Conference on Innovation and Technology in Computer Science Education ({ItiCSE})}}.
\newblock


\bibitem[Sarsa et~al\mbox{.}(2022)]%
        {DBLP:conf/icer/SarsaDH022}
\bibfield{author}{\bibinfo{person}{Sami Sarsa}, \bibinfo{person}{Paul Denny}, \bibinfo{person}{Arto Hellas}, {and} \bibinfo{person}{Juho Leinonen}.} \bibinfo{year}{2022}\natexlab{}.
\newblock \showarticletitle{{A}utomatic {G}eneration of {P}rogramming {E}xercises and {C}ode {E}xplanations {U}sing {L}arge {L}anguage {M}odels}. In \bibinfo{booktitle}{\emph{Proceedings of the Conference on International Computing Education Research ({ICER})}}.
\newblock


\bibitem[Smith et~al\mbox{.}(2024)]%
        {smith2024prompting}
\bibfield{author}{\bibinfo{person}{David~H. Smith}, \bibinfo{person}{Paul Denny}, {and} \bibinfo{person}{Max Fowler}.} \bibinfo{year}{2024}\natexlab{}.
\newblock \showarticletitle{{P}rompting for {C}omprehension: {E}xploring the {I}ntersection of {E}xplain in {P}lain {E}nglish {Q}uestions and {P}rompt {W}riting}. In \bibinfo{booktitle}{\emph{Proceedings of the Conference on Learning @ Scale (L@S)}}.
\newblock


\bibitem[Wei et~al\mbox{.}(2022)]%
        {DBLP:conf/nips/Wei0SBIXCLZ22}
\bibfield{author}{\bibinfo{person}{Jason Wei}, \bibinfo{person}{Xuezhi Wang}, \bibinfo{person}{Dale Schuurmans}, \bibinfo{person}{Maarten Bosma}, \bibinfo{person}{Brian Ichter}, \bibinfo{person}{Fei Xia}, \bibinfo{person}{Ed~H. Chi}, \bibinfo{person}{Quoc~V. Le}, {and} \bibinfo{person}{Denny Zhou}.} \bibinfo{year}{2022}\natexlab{}.
\newblock \showarticletitle{{C}hain-of-{T}hought {P}rompting {E}licits {R}easoning in {L}arge {L}anguage {M}odels}. In \bibinfo{booktitle}{\emph{Proceedings of the Annual Conference on Neural Information Processing Systems ({NeurIPS})}}.
\newblock


\bibitem[Whalley et~al\mbox{.}(2021)]%
        {whalley2021novice}
\bibfield{author}{\bibinfo{person}{Jacqueline Whalley}, \bibinfo{person}{Amber Settle}, {and} \bibinfo{person}{Andrew Luxton{-}Reilly}.} \bibinfo{year}{2021}\natexlab{}.
\newblock \showarticletitle{{N}ovice {R}eflections on {D}ebugging}. In \bibinfo{booktitle}{\emph{Proceedings of the Technical Symposium on Computer Science Education ({SIGCSE})}}.
\newblock


\end{thebibliography}

\end{document}